\documentclass[prl,twocolumn]{revtex4}
\usepackage{graphicx,amsmath,amssymb,amsxtra,times}
\renewcommand{\narrowtext}{\begin{multicols}{2} \global\columnwidth20.5pc}

\def\be{\begin{eqnarray}}
\def\ee{\end{eqnarray}}

\begin{document}
%\draft

%\title{Spin glassiness and power law scaling in the quasi-triangular spin-1/2 compound Cu$_{2(1-x)}$Zn$_{2x}$(OH)$_3$(C$_7$H$_{15}$COO)$\cdot $mH$_2$O}
\title{Spin glassiness and power law scaling in a quasi-triangular spin-1/2 compound}

\author{Jian Wu}
\author{Julia S. Wildeboer}
\author{Fletcher Werner}
\author{Alexander Seidel}
\author{Z. Nussinov}
\author{S. A. Solin}
\email[]{solin@wuphys.wustl.edu} 
\affiliation{Department of Physics and Center for Materials Innovation, Washington University, St. Louis, MO 63136, USA}

\date{\today}

\begin{abstract}
We present data on the magnetic properties of two classes of layered spin S=1/2 antiferromagnetic quasi-triangular lattice materials: Cu$_{2(1-x)}$Zn$_{2x}$(OH)$_3$NO$_3$($0\!<\!x\!<\!0.65$) 
and its long chain organic derivatives Cu$_{2(1-x)}$Zn$_{2x}$(OH)$_3$(C$_7$H$_{15}$COO)$\cdot$mH$_2$O($0\!<\!x\!<\!0.29$),
where non-magnetic Zn substitutes Cu isostructurally.
%These series of layered structure compounds constitute a substitutional magnetic system, in which 
%spin S=1/2 Cu$^{2+}$ ions and nonmagnetic Zn$^{2+}$ ions are arranged on a 2D quasi-triangular lattice. 
It is found that the long-chain compounds, even in a clean system in the absence of dilution, $x\!=\!0$,
show spin-glass behavior, as evidenced by DC and AC susceptibility, and by time dependent
magnetization measurements. 
A striking feature is the observation of a sharp crossover between two successive power law regimes in the 
DC susceptibility above the freezing temperature. 
%The high temperature power law is suggestive of % consistent with
 %a quantum Griffiths based scenario,
%with additional interesting structure at lower temperature.
Specific heat data are consistent with a conventional phase transition in the unintercalated compounds, and glassy behavior in the long chain compunds.
%will also be presented.
%Contents await final data
\end{abstract}

\maketitle
{\em Introduction.}
The study of frustrated quantum magnetism has defined new paradigms of condensed matter for many decades. When low dimensionality, geometrical frustration, and quantum fluctuations interplay, novel and interesting phase diagrams may result. Another driving factor of key importance is disorder. Much excitement has been generated by the recent discovery of a possible realization of the long sought ``spin liquid'' phase, in compounds such as herbertsmithite \cite{helton}. Another intricate state of matter, the spin glass, has been thoroughly studied in systems such as LiHo$_x$Y$_{1-x}$F$_4$\cite{reich}. Despite these efforts, the spin glass state (and, in particular, its realization in electronic systems) is still counted among the most enigmatic states of matter. Many questions remain unresolved. One such key question is whether spin glass type behavior requires explicit structural disorder of the host system, because of the short time scales generally associated with spin dynamics in clean systems, or whether it may be acquired spontaneously in a clean system (as in, e.g.,  ordinary glasses). Here, we present experimental results shedding light on this and other issues. 
 
 The systems under investigation are derivatives of copper-hydroxy-nitrate (CHN), Cu$_2$(OH)$_3$NO$_3$ in the presence of zinc doping.  The derivatives studied are (i) the diluted CHN system
 Cu$_{2(1-x)}$Zn$_{2x}$(OH)$_3$NO$_3$ and (ii) the long organic chain intercalates Cu$_{2(1-x)}$Zn$_{2x}$(OH)$_3$(C$_7$H$_15$COO)$\cdot$mH$_2$O. In the absence of doping, these layered hydroxides are nearly triangular S=1/2 antiferromagnets.
 Some properties of the undoped parent compounds have been studied earlier \cite{linder, epstein}.% for CHN \cite{linder} and the long chain derivative \cite{epstein}. 
 Evidence of glassiness has been reported for the $x\!=\!0$ long chain compounds, 
 which were argued to be clean systems described by
 frustrated short range magnetic interactions without much disorder\cite{epstein}. 
 
 In recent years, chemical doping has proven to be an invaluable tool in accessing an extended phase
 diagram of transition metal oxides, in particular for detecting otherwise inaccessible transitions.
 Examples are the 
 high temperature superconducting cuprates and pnictides, and various heavy fermion compounds \cite{bednorz, pnictides, fisk,zohar},
 some of which also exhibit glassiness in some parameter regime.
 In this work, we apply this tool to study the CHN compound and its long chain derivative.
 In doing so, we discover a remarkably robust behavior of the long chain compounds 
 which manifests itself through two distinct regimes of power law
 scaling in the temperature dependence of the DC susceptibility, separated by a sharp crossover.
 Moreover, we also identify this behavior in the undoped parent material. We find that Zn doping
 enhances this effect  and 
 shows it to be a robust feature of the physics of the
 glassy long chain compounds.
Furthermore, we present significant new evidence for the glassiness of both the doped and undoped
long chain compounds. The ability to control the Zn concentration sheds further light on the relation
between structural disorder and
the glassiness of these compounds.

 %In an effort

  %to control the amount of structural disorder in the system,
 %we have found a technique\cite{wu10} of uniformly tuning the concentration of magnetic Cu$^{2+}$ 
 %ions without phase separation nor notable distortion of the lattice structure. We present new evidence for spin glass physics
 %in both undoped and Zn-doped long chain samples and monitor the glassy behavior in DC and AC susceptibility as well as time dependent
 %magnetization measurements as a function of Zn concentration. We further identify a sharp crossover between two power law regimes in the
 %DC susceptibility that is universally present in all samples. This scaling behavior may hint at the presence of 
 %Griffiths type physics and/or quantum critical points ``shielded'' by glassy order at low temperatures, making the CHN family an ideal
 %class of systems to study the interplay of glassiness and quantum critical points.
 
 \begin{figure}
 \includegraphics[width=9cm]{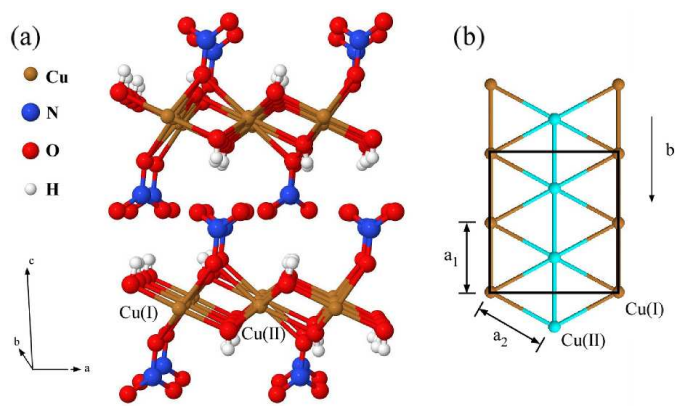}
 \caption{\label{structure}(Color online). Structure of the CHN compound. a) perspective view of the basal unit.  b) c-axis view of the Cu layer and its planar unit cell (dark rectangle).  $a_1 = 3.03$\AA,  $a_2 = 
3.17$\AA. 
}
 \end{figure}
 
 {\em Structure and preparation.}
 The structure of the CHN compound is shown in Fig. \ref{structure}.
 The unit cell contains two inequivalent Cu$^{2+}$ ions forming nearly triangular layers.
 The preparation of the undoped samples follows Refs. \onlinecite{linder} for CHN and \onlinecite{epstein} for the long chain
 compounds. The method for preparing the Zn-doped samples has been reported earlier \cite{wu10}.
 Powder X-ray diffraction measurements have been carried out \cite{wu10}.
 For both long chain intercalates and diluted CHN samples, Zn was found to replace Cu leading to a series of isostructural doped compounds.
 The long chain samples are characterized by
 pronounced $(00l)$ reflections, as is typical for layered structure compounds with large interlayer distances. 
 From this the interlayer distance was obtained to be $24.2$\AA, consistent with Ref. \onlinecite{epstein}.
 This distance was also found to be independent of Zn doping concentration.  A detectable ZnO impurity phase was 
only observed in the x = 0.65 sample.

   \begin{figure}[t]
 \includegraphics[width=9cm]{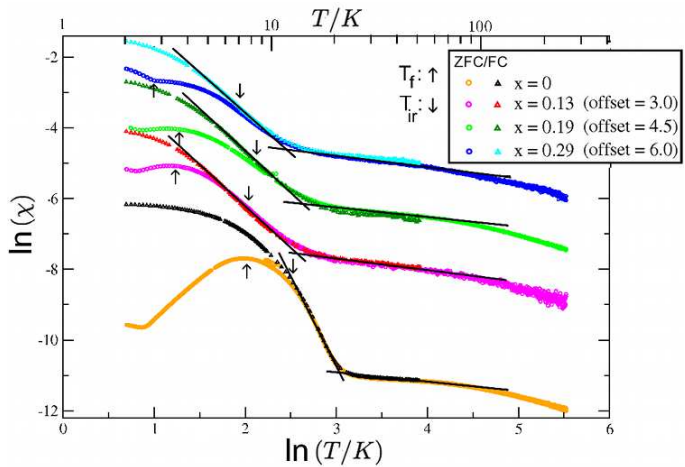}
 \caption{\label{DC} (Color online). 
$\ln(\chi)$ vs. $\ln(T)$ for DC susceptibility (long chain samples). Both field cooled and zero field cooled data are shown,
with different offsets for different dopings $x$ as indicated.
The freezing temperature $T_f$ and irreversibilty temperature $T_{ir}$ are indicated by arrows.
Linear fits identify various temperature regimes where $\chi(T)$ is well described by power laws.
In the lower temperature regimes, field cooled data was used for the fit. However, the zero field
cooled data is seen to follow similar and in some cases almost identical power laws .
}
 \end{figure}

   \begin{figure}[t]
 \includegraphics[width=8cm]{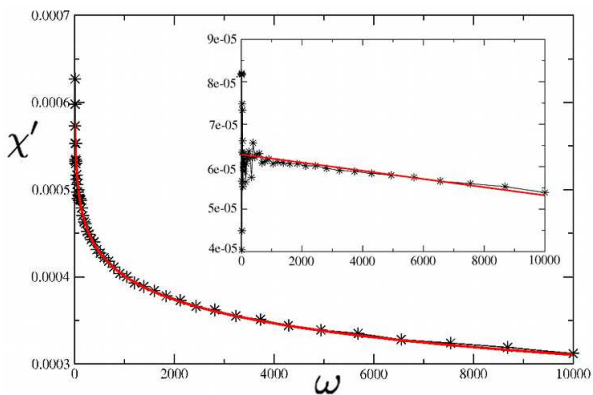}
 \caption{\label{AC} (Color online). 
AC susceptibility of $x=0$ sample at $T=10K$.
The real part is fitted to the functional form $\chi'=\chi_0-a\ln(\omega)$.
The linear fit of the imaginary part (inset) approaches $\pi a/2$ at $\omega=0$
to within $2\%$ accuracy.
}
 \end{figure}

{\em DC and AC susceptibility.}
DC susceptibility data for the diluted CHN system has been reported in Ref. \onlinecite{wu10}.
Both the Curie-temperature and the N\'eel temperature vary with Zn concentration, but are on the order of 10K
in agreement with density functional values for the exchange couplings\cite{dft}.
The value of the exchange interaction parameter is known to be sensitive to the Cu-OH-Cu bond angle, with both 
ferromagnetic (FM) and anti-ferromagnetic (AFM) values being possible. In CHN the angle is close
to the FM-AFM crossover\cite{linder} rendering the material weakly anti-ferromagnetic.
Interestingly, with increased Zn concentration the Curie constant was found to change sign and
become ferromagnetic\cite{wu10}. There was, however, no FM phase seen at low temperature
and a complete analysis of this behavior would likely need to take into account Dzyaloshinskii-Moriya
interactions, which are allowed due to the relatively low symmetry at the Cu site\cite{epstein}.

We will now focus on the long chain susceptibility data shown in Fig. \ref{DC}. A clear difference between
the zero field cooled (ZFC) and field cooled (FC) data below a temperature  $T_{ir}\sim 6-12$K
is a first indication of the glassiness of the long chain samples. 
No similarly large effect has been seen in the diluted CHN samples\cite{wu10}.
It is worth pointing out that in the long chain case, the ZFC/FC difference actually diminishes 
at finite  Zn concentration compared to the undoped parent compound. This may indicate that the glassiness of the system
is indeed driven by frustration rather than structural disorder. Similarly, the freezing temperature
$T_f$ as determined by the peak in the ZFC susceptibility is around 8K for the $x\!=\!0$ sample, and
is about a factor of 2 lower for the Zn-doped samples. 

A remarkable feature seen in the
log-log plots of Fig. \ref{DC} is the presence of a sharp crossover between two 
 distinct power law regimes above $T_f$.
 The first of these regimes corresponds to a temperature window typically between 5K and 12K,
 where $\chi\propto T^{-a}$ with different exponents $a \!\simeq\! 5$ for x=0 and $a\simeq 2$ for $x>0$ .
 In a second regime  
 at higher temperatures roughly between 12K and 90K, $\chi\propto T^{-b}$ with $b \simeq 0.3$ for all samples,
 below a Curie tail with $\chi\propto T^{-1}$ for $T\!>\!90K$. 
 This behavior is slightly more pronounced in the $x>0$ samples which all have very similar exponents, but is clearly identifiable also
 in the $x=0$ sample.
 Hence the sharp crossover between two different power law regimes as a precursor to the spin glass phase appears to be a robust feature of the physics of the long chain samples. 
We note that FC and ZFC data are indistinguishable in the higher temperature regime. In the 
lower temperature regime, the FC data have been used for the linear fits in Fig. \ref{DC}.
However, the ZFC data display power law behavior with similar exponents even there.

The frequency dependence of the AC susceptibility of the x = 0 long chain sample is shown in  Fig. \ref{AC}. The real part $\chi'$ was found to be extremely well described by the logarithmic functional form $\chi'=\chi_0-a\ln(\omega)$  at low frequencies $<10^4$ Hz, a standard %textbook 
behavior\cite{hertz} for spin glasses.
This is 
  in good agreement with the imaginary part (inset), whose low frequency 
  limit is a non-zero constant that agrees well with
  $\pi a/2$ as determined from the real part via Kramers-Kronig relations. 
  These findings provide strong new evidence for spin-glass physics in the
  x = 0 long chain sample, even though this system seems to have no apparent structural disorder.

   \begin{figure}[t]
 \includegraphics[width=6.5cm]{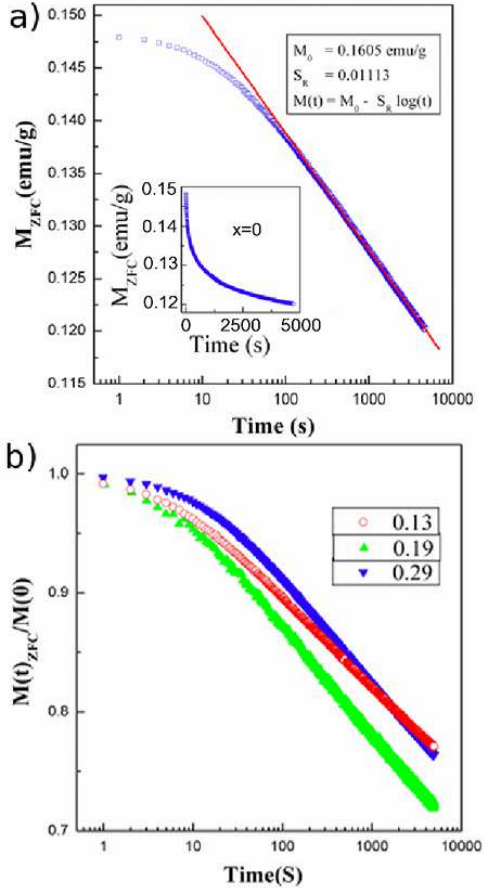}
 \caption{\label{mag} (Color online). Isothermal remanent magnetization  $M_{ZFC} (t)$ vs. $\log(t)$ at 5000 Oe. a) $x = 0$. Inset: The normal plot of $M_{ZFC}(t)$ for $x = 0$. b) The scaled remanent magnetization $M_{ZFC} (t)/ M(0)$ at 5000 Oe for $x =0.13$, $0.19$ and $0.29$.
}
 \end{figure}

{\em Time dependent magnetization.}
The time evolution of the isothermal remanent magnetization $M_{ZFC}(t)$ of the long chain samples has been studied in the following steps.  The sample was first cooled in zero field to 2 K (well below the freezing temperature) and then kept in a magnetic field, H, for 10 minutes.  After switching off the applied field, we measured the remanent magnetization $M_{ZFC}(t)$ as a function of time for 5000 seconds. The experiment was repeated in 5 different fields (50 Oe, 500  Oe, 5000 Oe, 10000 Oe and 20000 Oe). The time dependence of $M_{ZFC}(t)$ of the x = 0 sample is plotted in the inset of  Fig. \ref{mag} a), in which we observed a long time slow non-exponential relaxation.  For $t  \ge100$s, $M_{ZFC}(t)$ can be described by
$M_{ZFC}(t)=M_0-S_R \log(t)$,
in which $M_0$ is a constant and the coefficient $S_R$ is the Òmagnetic viscosityÓ.  This logarithmic decay law is a characteristic feature of spin glass like systems. In order to better compare the relaxation behaviors of other doped samples, a scaled $M_{ZFC}(t)/M(0)$ quantity is plotted in Fig. \ref{mag}b) instead of the absolute values. At 5000 seconds, their $M_{ZFC}(t)$ values slowly decay to $70-80\%$ of the initial values $M(0)$.  Their long time decay behaviors all follow the logarithmic form with different coefficients $S_R$.

   \begin{figure}[t]
 \includegraphics[width=8cm]{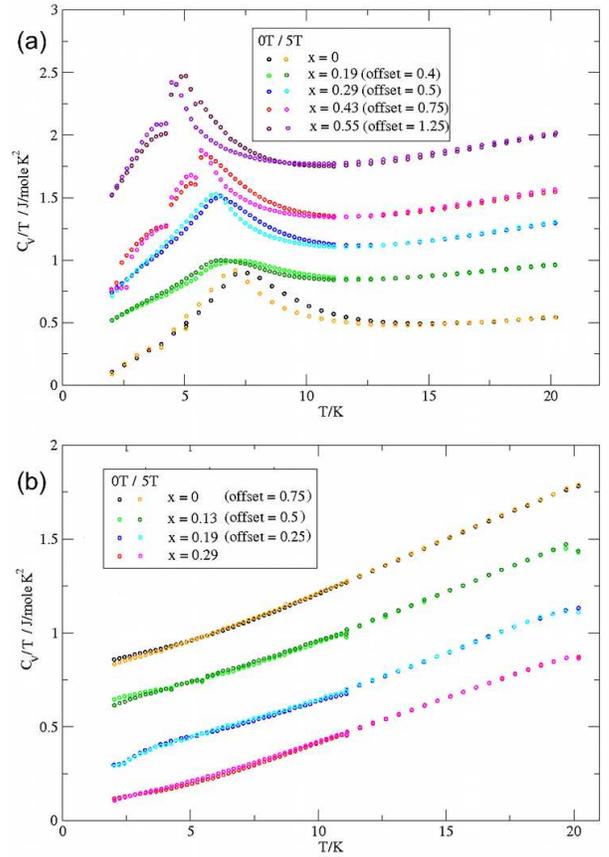}
 \caption{\label{spec} (Color online). 
 Specific heat divided by temperature, for diluted CHN (a) and long chain (b) samples, with and without
 an applied magnetic field of $B=5T$. Data for different Zn concentration $x$ are mutually offset for clarity.
}
 \end{figure}

{\em Specific heat.}
Fig. \ref{spec} shows $C/T$ vs. temperature $T$, where $C$ is the specific heat, for diluted CHN samples (a) and long chain samples (b).
Sharp peaks  in the diluted CHN data indicate a large amount of entropy loss during the antiferromagnetic phase transitions.
This  supports a picture based on the existence of conventional long-range 
magnetic order in the diluted CHN samples. 
The peak position of C/T
decreases as the Zn doping increases, which is similar to the doping dependence of the DC susceptibility peaks
\cite{wu10}.
We should also note that in the temperature range $2$K$<\!T\!<\! 4$K, the specific heat has a roughly $T^2$ behavior.
This is also seen up to $T\simeq 20K$ or more in the long chain data, Fig. \ref{spec}b), as noted for $x=0$ in Ref. \onlinecite{epstein}.
We caution that
since there are no proper nonmagnetic materials with an analogous lattice structure available, the pure magnetic contribution cannot be easily
separated. To shed more light on this issue, we compared the specific heat in zero field to that in a field of $B=5T$,
where the Zeeman energy is comparable to the magnetic exchange interaction (see above).
This has a pronounced effect on the peak structure in the diluted CHN samples, giving further evidence of the 
magnetic origin of this peak. On the other hand, the effect of the magnetic field on the overall specific heat is very small in the
long chain samples. The small difference between $B=0$ and $B=5T$ is barely visible in Fig. \ref{spec}b), 
and is found to be of order $0.1$J/mol K. 
This small magnetic field dependence indicates that lattice degrees of freedom dominate the
overall specific heat. 
One may attribute this dominance to the large unit cell of the system,
especially for the long chain samples, and to the enhanced phonon density
of states at low energies due to the approximately two-dimensional character of the system.
Furthermore, the magnetic peak (and the associated field dependence)
is seen to be absent in the long chain samples.
 This is also a common feature seen in many spin glass like systems\cite{hertz}, 
 due to the gradual freezing of spins or spin clusters over a large temperature range. 
 In particular, if spins are locked into large clusters even above $T_f$,
 this may provide an additional explanation
 why magnetic degrees of freedom do not release much entropy at low temperatures,
 and contribute relatively little to the low-$T$ specific heat.
 From the weak magnetic field dependence, we conclude that the approximate $T^2$-behavior
 of the low-$T$ specific heat
  is chiefly due to phonons.
 This is a direct consequence of the quasi-two-dimensional geometry of the system.
 However, a close inspection of the graphs in Fig.\ref{spec} b) shows that the low-$T$ limit of $C/T$ is non-zero,
 hence implying a weak linear contribution to the specific heat. This is the expected low-$T$ contribution 
 of a spin glass (or structural glass) phase\cite{hertz, binder}.

{\em Discussion.}
The CHN family of compounds allows a controlled study of the role of non-magnetic impurities
in a layered frustrated spin-1/2 compound with glassy behavior. 
We have given further evidence for the presence of a spin glass phase in the ``clean" ($x=0$)
long chain parent compound, and tracked the observed phenomenology as a function
of Zn concentration $x$.
The $x=0$ limit in the absence
of Zn substitution was seen to be the case where the glassy features were most pronounced.
This may strengthen earlier claims according to which the spin glass phase is driven by
frustration rather than disorder. 
It is likely that anisotropy introduced by Dzyaloshinskii-Moriya 
interactions also plays a role in the observed behavior\cite{epstein,collins}, especially in view
of the weakness of the energy scale associated with exchange interactions.
We caution that  Dzyaloshinskii-Moriya  terms are particularly sensitive to the symmetry of the local environment,
and thus could be more affected by disorder introduced by the organic chains in the long chain samples.
Detailed model calculations on the required strength of these terms and the amount of disorder necessary
to reproduce the observed behavior are left for future studies.

A remarkable effect was found to emerge at temperatures above $T_f$ in the 
form
of two successive power law regimes in the DC susceptibility with sharp crossover.
This effect was furthermore seen to be robust against Zn doping, becoming rather more
pronounced in the $x>0$ samples. The occurrence of power laws above a transition into 
a glassy state is somewhat reminiscent of the picture developed in Ref.\onlinecite{vlad}.
There it was argued that a quantum Griffiths phase\cite{griffiths} may be unstable to the formation
of a cluster glass phase at low temperatures. Above the transition temperature, the quantum
Griffiths behavior is still expected to be seen, which leads to the observation of power laws
in thermodynamic quantities as determined in Ref. \onlinecite{vojta}. Specifically,
the susceptibility is predicted to be of the form $\chi\propto T^{\alpha-1}$ with 
$\alpha>0$.
%$\alpha=d/z'$, where $d$ is the
%dimensionality of the system, and $z'$ is a ``dynamical exponent'' describing the distribution of clusters\cite{young}.
It is feasible that the role of long range RKKY interactions considered  in Ref. \onlinecite{vlad} is played by
dipole-dipole interactions in the present case.
We note, however, that the value of $\alpha$ in this scenario is not universal, but is expected
to depend on doping\cite{Ce}.  
In contrast, the power laws observed here are fairly insensitive to doping
which is more suggestive of universal exponents.
%that the low temperature instability of the Griffiths phase predicted in Ref. \onlinecite{vlad} is due 
%to long range power law interactions, which in Ref. \onlinecite{vlad} were taken to be of RKKY form.
%However, it is feasible that a similar instability could be triggered by dipole-dipole interactions
%in the class of non-itinerant magnets studied here. 
%We note, however, that
Furthermore, 
the quantum Griffiths
scenario can at present only 
explain the occurrence of a single power law with exponent less than 1,
and offers no direct explanation for the power law behavior seen at low temperature.
%This may well describe
%the higher temperature power law regime in our study, and also
%seems to capture
%the behavior of the Ce based compound recently studied in Ref. \onlinecite{Ce}. 
To our knowledge,
a crossover between different power laws as seen in the long chain CHN compounds
has not been previously observed or predicted, and may well be the hallmark of new physics.
A detailed understanding of this behavior 
and the search for possible incarnations in other systems
remain an interesting challenge for future work.\\
\begin{acknowledgements}
We would like to thank the Center for Materials Innovation for support of this work.
AS would like to acknowledge support
by the National Science Foundation under NSF Grant No. DMR-0907793,
as well as insightful discussions with V. {Dobrosavljevi\ifmmode \acute{c}\else \'{c}\fi{}}.
\end{acknowledgements}
%\vspace{-9mm}
%\bibliography{CHN}
%merlin.mbs 2010-03-15 4.21a (PWD, AO, DPC)

%These may be attributed to cluster formation. 

%Specific heat data will also be presented.
 
\end{document}